\documentclass[a4paper,11pt]{article}
\usepackage{pos}
\usepackage{wrapfig}
\usepackage{enumitem}
\usepackage{makecell} 

\usepackage{colortbl}
\usepackage{xspace}
\usepackage{xcolor}

\usepackage[symbol]{footmisc}

\definecolor{codegreen}{rgb}{0,0.6,0}
\definecolor{codegray}{rgb}{0.5,0.5,0.5}
\definecolor{codepurple}{rgb}{0.58,0,0.82}
\definecolor{backcolour}{rgb}{0.95,0.95,0.94}
\definecolor{SWGOOrange}{rgb}{0.906,0.478,0.200}
\definecolor{SWGOOrangeLight}{rgb}{1.0,0.678,0.350}

\renewcommand{\thefootnote}{\fnsymbol{footnote}}

\title{The Southern Wide-field Gamma-ray Observatory}
\ShortTitle{The Southern Wide-field Gamma-ray Observatory}

\author*[a,b]{Ruben Concei\c{c}\~ao}
\onbehalf{for the SWGO Collaboration\footnote[3]{For the complete list of the SWGO Collaboration and acknowledgments, please visit: \url{https://www.swgo.org/SWGOWiki/lib/exe/fetch.php?media=wiki:swgo_collaboration_icrc23.pdf}}}

\affiliation[a]{LIP - Laborat\'orio de Instrumenta\c{c}\~ao e F\'isica Experimental de Part\'iculas, Lisbon, Portugal}
\affiliation[b]{Departamento de F\'isica, Instituto Superior T\'{e}cnico, Universidade de Lisboa, Lisbon, Portugal}

\emailAdd{swgo\_spokespersons@swgo.org}

\abstract{The Southern Wide-field Gamma-ray Observatory (SWGO) is an R\&D project to plan and design the next observatory to detect gamma rays in the Southern hemisphere. The experiment, planned to be placed at an altitude greater than 4400 m, is primarily based on water Cherenkov detectors units and is expected to measure gamma rays from a few hundred GeV up to the PeV scale. SWGO will complement CTA and the existing ground-based particle detectors of the Northern Hemisphere, namely HAWC and LHAASO, having a rich science programme.
The collaboration is highly invested in evaluating different detector and array configurations, prototyping, and site search. In this presentation, I shall present an overview of the project's activities, achievements and future plans.}

\ConferenceLogo{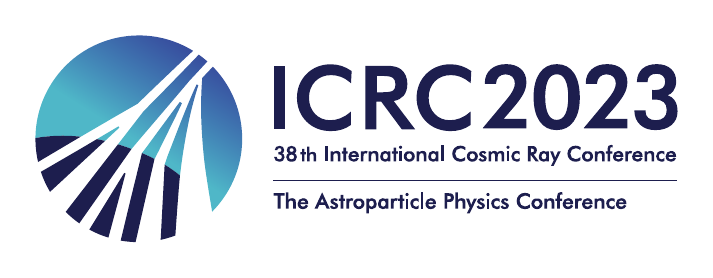}

\FullConference{%
38th International Cosmic Ray Conference (ICRC2023)\\
26 July -- 3 August, 2023\\
Nagoya, Japan}

\begin{document}
\maketitle

\section{The SWGO collaboration}
The Southern Wide-field Gamma-ray Observatory (SWGO) collaboration was established in 2019 with the collective goal of creating a ground-particle-based gamma-ray observatory to survey and monitor the Southern Hemisphere sky. It emerged as a merger of precursor projects, such as SGSO~\cite{SGSO} and LATTES~\cite{LATTES}. SWGO is an international collaboration involving 14 countries: Argentina, Brazil, Chile, China, Croatia, the Czech Republic, Germany, Italy, Mexico, Peru, Portugal, South Korea, the United Kingdom, and the United States of America. These partner institutes, totalling 64, have showcased their commitment to the development of SWGO by endorsing the "Statement of Interest". Notably, the strong involvement of South American countries is evident, as the observatory is envisioned to be situated at high altitudes in the Andes. Furthermore, the collaboration receives support from scientists in 10 additional countries.

The SWGO collaboration capitalizes on the expertise gained from previous successful ventures in both extensive air shower ground arrays, such as HAWC, LHAASO, the Pierre Auger Observatory, and IceCube/IceTop, and imaging Cherenkov telescopes, including MAGIC and HESS.

\section{R\&D phase plan}
SWGO is currently in the research and development phase, with a focus on completing the Conceptual Design Report for the observatory. This involves crucial tasks such as establishing a baseline design, selecting an ideal site, and defining benchmark science cases. Despite slight delays caused by the COVID pandemic, the R\&D phase is expected to conclude by the end of 2024, followed by a Preparatory Phase centered around engineering finalization, project management, and resource identification. The collaboration operates according to a well-established R\&D plan, outlined by the milestones presented in Table~\ref{tab:Milestones}.

Throughout the R\&D phase, SWGO followed a systematic approach to optimize scientific performance within a fixed cost framework. It involved defining and evaluating various options for each detector element, based on predefined Benchmark Science Cases. A Reference Design was established as a benchmark, and candidate configurations developed to cover a range of science optimizations with costs equivalent to the Reference Design. Monte Carlo simulations using a Reference Analysis chain assess the response of each candidate configuration to gamma-rays and background events. Through site evaluations and comparisons against the Benchmark Science Cases, a preferred site, configuration, and design options are collectively chosen, while contingency plans are also considered. The chosen configuration undergoes further refinement based on the selected site and remaining technical considerations, leading to the development of a Conceptual Design Report that outlines the Baseline configuration, expected performance, and construction and operation concepts.

\begin{table}[h]
\centering
\begin{tabular}{ll}
\Xhline{4\arrayrulewidth}
\rowcolor{SWGOOrange}
\textcolor{white}{M1} & \textcolor{white}{R\&D Phase Plan Established}   \\
\rowcolor{SWGOOrange}
\textcolor{white}{M2} & \textcolor{white}{Science Benchmarks Defined}                        \\
\rowcolor{SWGOOrange}
\textcolor{white}{M3} & \textcolor{white}{Reference Configuration \& Options Defined}        \\
\rowcolor{SWGOOrange}
\textcolor{white}{M4} & \textcolor{white}{Site Shortlist Complete}                           \\
\rowcolor{SWGOOrange}
\textcolor{white}{M5} & \textcolor{white}{Candidate Configurations Defined}                  \\
M6 & Performance of Candidate Configurations Evaluated \\
M7 & Preferred Site Identified                         \\
M8 & Design Finalised                                  \\
M9 & Conceptual Design Report Complete                 \\
\Xhline{2\arrayrulewidth}
\end{tabular}
\caption{The Milestones of the current SWGO research and development phase. The orange (filled) cells correspond to the Milestones completed.}
\label{tab:Milestones}
\end{table}

SWGO's R\&D activities are organized into five major working groups (WGs): Science, Analysis and Simulation, Detector, Site, and Outreach and Communication. Each WG is led by 2-3 coordinators who hold regular meetings to drive progress. The collaboration also benefits from an advisory group comprising experienced individuals in the field. The main decision-making body of the collaboration is the Steering Committee, consisting of representatives from the 14 different countries listed above.

\section{Science Goals}

The possibilities for an observatory such as SWGO are enormous raging from astrophysics to fundamental particle physics~\cite{GC_Poster,DM_Poster,BSM_SWGO,CosmicRays_Poster}. In SWGO, core science cases have been carefully defined to provide guidance for the R\&D studies and to serve as benchmarks for evaluating various options and trade-offs in the final observatory design.
Table~\ref{tab:benchmarks} presents the six core science cases that SWGO is actively pursuing, along with their main design drivers for the experiment and the corresponding benchmarks under consideration. These benchmarks represent a minimum set of science goals that encompass the complete range of performance requirements for the observatory. Utilizing quantitative benchmarks, a thorough comparison will be conducted to select a set of candidate configurations for the array, which are currently being studied.

\begin{table*}[h]
    \centering
    {\small
    \begin{tabular}{|p{4.3cm}|p{4.3cm}|p{5cm}|}
        \hline
        \rowcolor{SWGOOrange}
        \normalsize{\textbf{\textcolor{white}{Core Science Case}}} & \normalsize{\textbf{\textcolor{white}{Design Drivers}}} & \normalsize{\textbf{\textcolor{white}{Benchmark Description}}} \\
        \hline \hline

        \textbf{Transient Sources: \newline Gamma-ray Bursts} &
        Low-energy \newline Site altitude &
        Min. time for $5 \sigma$ detection \newline
        F(100$\,$GeV)$ = 10^{-8}\,{\rm erg\,cm^{-2}\,s^{1}}$ \\
        \hline

        \textbf{Galactic Accelerators: \newline PeVatron Sources} &
        High-energy sensitivity \newline Energy resolution &
        Maximum exp-cutoff energy detectable $95\%$ CL in 5 years for: \newline
        F(1$\,$TeV)$ = 5\,$mCrab, index$=-2.3$ \\
        \hline

        \textbf{Galactic Accelerators: \newline PWNe and TeV Halos} & Extended source sensitivity \newline Angular resolution & Max. angular extension detected at $5\,\sigma$ in 5-yr integration for: \newline
        F(>1$\,$TeV)$ = 5\times10^{-13}\,{\rm TeV\,cm^{-2}\,s^{1}}$ \\
        \hline

        \textbf{Diffuse Emission: \newline Fermi Bubbles} &
        Background rejection &
        Minimum diffuse cosmic-ray residual background level.\newline
        Threshold: $< 10^{-4}$ level at $1\,$TeV. \\
        \hline

        \textbf{Fundamental Physics: \newline Dark Matter from GC Halo} &
        Mid-range energy sensitivity \newline Site latitude & Max. energy for $b\bar{b}$ thermal relic cross-section at $95\%$ CL in 5-yr, for Einasto profile. \\
        \hline

        \textbf{Cosmic-rays: \newline Mass-resolved dipole \newline Multipole anisotropy} & Muon counting capability & Max. dipole energy at $10^{-3}$ level. Log-mass resolution at $1\,$PeV $-$ goal is $A = {1,4,14,56}$; Maximum multipole scale $> 0.1\,$PeV. \\
        \hline

    \end{tabular}
    }
    \caption{SWGO Science Benchmarks and associated design drivers. Flux sensitivities are all calculated for 5 years, and the quoted energy threshold is defined at near-peak detection effective area, to provide a source-independent reference.}
    \label{tab:benchmarks}
\end{table*}

The final design of the observatory will inevitably involve a careful balance between the physics reach, technological feasibility, and cost considerations. Nevertheless, the science core cases provide valuable insights that allow for certain performance constraints to be established for the observatory. As an example, the objective of observing transient sources imposes a requirement for a low energy threshold, which directly influences the choice of the future site altitude. Currently, an altitude above $4.4\,$km a.s.l. is being considered. The search for galactic accelerators demands an energy resolution better than $\mathcal{O}(30\%)$ across the energy range of $1-100\,$TeV, as well as an angular resolution of approximately $0.15^\circ$. Additionally, having an excellent capability for gamma/hadron discrimination and sensitivity to cosmic-ray mass composition groups is highly desirable. Consequently, the design of the water-Cherenkov Detector (WCD) units should be optimized to accurately determine the muon content of extensive air showers.

\section{Simulation framework}

Within the SWGO R\&D efforts, the assumption is that the detector station units will primarily consist of water Cherenkov detectors (WCDs). WCDs have demonstrated their reliability in detecting the secondary particles of shower events, providing calorimetric information on the ground footprint, and even detecting muons~\cite{LHAASO_muon, Auger_muons}. They have been successfully employed in experiments such as HAWC, LHAASO, and the Pierre Auger Observatory.

SWGO has made significant progress in developing its own comprehensive simulation and event reconstruction framework, enabling the exploration and equitable comparison of various detector concepts and array layouts.

The simulation framework comprises four integrated major structures: CORSIKA, AERIE, SWGO-RECO, and PySWGO. CORSIKA~\cite{CORSIKA} is employed to simulate the extensive air showers (EAS) generated by gamma or cosmic rays interacting with the atmosphere. The resulting particles at the ground are then fed into AERIE, which is the simulation framework inherited from HAWC. Leveraging their accumulated experience in detecting high-energy gamma rays with EAS arrays, AERIE has been adapted to incorporate modularity, allowing for the simulation of different detector concepts and array layouts. The simulated data is analyzed using SWGO-RECO, an application within AERIE that employs various reconstruction modules to estimate shower characteristics, including energy, direction, and core position~\cite{Template_Rec}.

Moreover, to enhance the capabilities of SWGO, a Python3-based layer has been developed as a complementary addition to the existing AERIE framework, which is based on C++ and Python2. This higher-level analysis layer plays a crucial role in generating Instrument Response Functions (IRFs), enabling performance comparisons between different detectors and serving as vital inputs for assessing the science case requirements. Being written in Python3, PySWGO offers also the possibility of using modern advanced tools in SWGO such as machine learning algorithms.

SWGO strives to push the boundaries by redesigning the detector concepts to achieve enhanced gamma/hadron discrimination power and efficient identification of EAS muons~\cite{Sb_poster}. One of the ideas being explored involves constructing WCDs with two chambers~\cite{DLWCD}, where the bottom chamber would be primarily sensitive to muons. Alternatively, small WCD units equipped with multiple photo-sensors are being investigated, allowing for muon identification through machine learning techniques~\cite{Mercedes}. Furthermore, new shower observables, such as azimuthal asymmetries of the shower footprint~\cite{LCm}, are being developed and tested, demonstrating promising results particularly at the highest energies.

\section{Detector Options and Site Candidates}

\begin{figure}[h]
  \centering
  \includegraphics[width=0.90\textwidth]{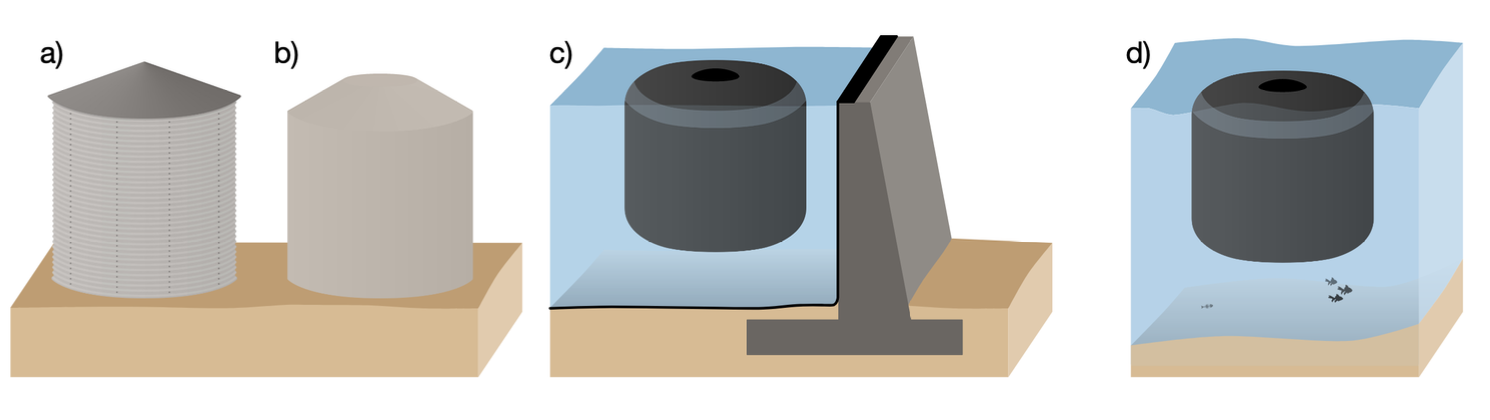}
  \caption{Detector concepts under study: cylindrical tanks constructed from (a) corrugated steel sheets or (b) roto-moulded HDPE; (c) open pond with floating bladder; (d) natural lake with floating bladder.
    \label{fig:DetectorConcepts}}
\end{figure}

To ensure the optimal design for the water Cherenkov detector units, SWGO is exploring different detector technologies~\cite{detector_poster}, including tanks, ponds, and lakes (see Fig.~\ref{fig:DetectorConcepts}). The first option involves using individual tanks, which would be mechanically separated and independently deployed. These tanks could be constructed with light-tight liners made of either roto-moulded plastic (similar to the Auger experiment) or steel (similar to HAWC).

The second option under consideration is the use of multiple large artificial water volumes, referred to as ponds. These ponds would incorporate retaining walls and optical separation between the units, resembling the setup of the Water Cherenkov Detector (WCD) used in LHAASO.

Option d) displayed in Fig.~\ref{fig:DetectorConcepts} involves deploying detector unit bladders filled with pure water directly into a natural lake~\cite{lake}. This approach entails placing the detectors in bladders and submerging them within a suitable lake.

Each of these options requires comprehensive evaluation in terms of cost, technical feasibility, and consideration of environmental and detector-related risks.

Apart from gathering information, several prototypes are under construction~\cite{prototypes,mercedes2,prototypes2}. These prototypes are going to be evaluated both in laboratory and in-site high-altitude conditions, demonstrating the option reliability.

Significant efforts are also being dedicated to the development of Data Acquisition Systems (DAQ) and the selection of suitable photo-sensors within the SWGO project.

SWGO is also invested in selecting an adequate site to build the experiment~\cite{site_poster}. A comprehensive data collection process has been conducted for the different candidate sites.
This valuable information has been gathered through collaborative efforts with members from the hosting countries as well as through dedicated site visits conducted by SWGO collaboration members.
Various factors have been taken into consideration, including altitude, local topology, environmental conditions, site access, transport costs, as well as the availability and cost of essential resources such as water, power, and network connectivity. In order to gather detailed information about the site conditions, an autonomous station specifically designed for environmental characterisation has been developed and deployed at each candidate site~\cite{env_poster}.

\begin{table*}[h]
    \centering
    {\small
    \begin{tabular}{|p{2.2cm}|p{2.2cm}|p{2.2cm}|p{2.2cm}|p{2.2cm}|}
        \hline
        \rowcolor{SWGOOrange}
        \normalsize{\textbf{\textcolor{white}{Country}}} & \normalsize{\textbf{\textcolor{white}{Site Name}}} &
        \normalsize{\textbf{\textcolor{white}{Altitude \newline [m a.s.l.]}}} &
        \normalsize{\textbf{\textcolor{white}{Latitude}}} &
        \normalsize{\textbf{\textcolor{white}{Notes}}} \\
        \hline \hline

        \textbf{Argentina} & Alto Tocomar  & 4,430 & 24.19 S & \\
        & Cerro Vecar & 4,800 & 24.19 S & Primary \\
        \hline


        \textbf{Chile} & Pajonales & 4,600 & 22.57 S & \\
        & Pampa La Bola & 4,770 & 22.25 S & Primary \\
        \hline

        \textbf{Peru} & Imata & 4,450 & 15.50 S & \\
        & Sibinacocha & 4,900 & 13.51 S & Lake site \\
        & Yanque & 4,800 & 15.44 S & Primary \\
        \hline

    \end{tabular}
    }
    \caption{SWGO candidate sites.}
    \label{tab:benchmarks}
\end{table*}








\begin{figure}[h]
  \centering
  \includegraphics[width=0.8\textwidth]{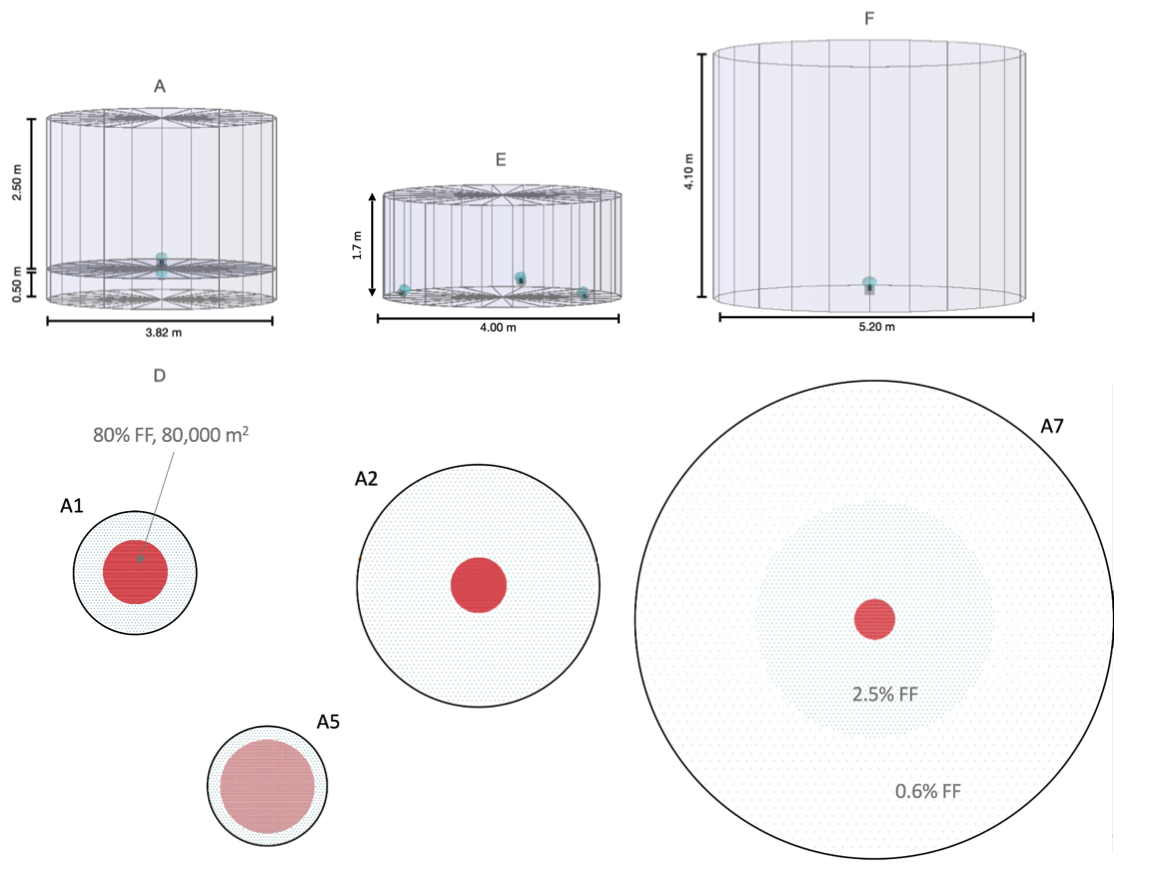}
  \caption{Top: Examples of the six water-Cherenkov detector unit configurations currently being studied for SWGO. Bottom: Illustrative examples showcasing the seven array configuration options currently under investigation.
    \label{fig:SummaryDetArrayM5}}
\end{figure}

\section{Current status and future plans}

The selection of a detector concept, array layout, and site for SWGO involves intricate correlations and represents a multifaceted problem. Over the past years, the collaboration has dedicated significant efforts to address these challenges, factoring in various considerations and gathering crucial information on detector technologies and candidate site conditions. Notably, the development of a simulation framework has played a pivotal role in exploring the available phase space, allowing for investigations into different detector concepts and array layout configurations. As a result, SWGO has entered a particularly exciting phase, where the collaboration is actively comparing diverse ideas and approaches to identify the best cost-effective solution for constructing the next-generation gamma-ray observatory.

The collaboration has currently completed a significant number of simulations for the detector and array configurations, known as Milestone 5. Figure~\ref{fig:SummaryDetArrayM5} showcases several examples of the 14 detector and array layout configurations being assessed. These configurations have been chosen to investigate key design elements and array configurations while maintaining a consistent cost framework. Parameters such as station dimensions, number and size of the photo-sensors, and the balance between compact (for lower energies) and sparse array (for higher energies) are being thoroughly examined. This ongoing exercise, set to be completed by the fall of this year, will provide valuable insights into identifying the most favourable options to be considered.

\begin{figure}[!h]
  \centering
  \includegraphics[width=0.80\textwidth]{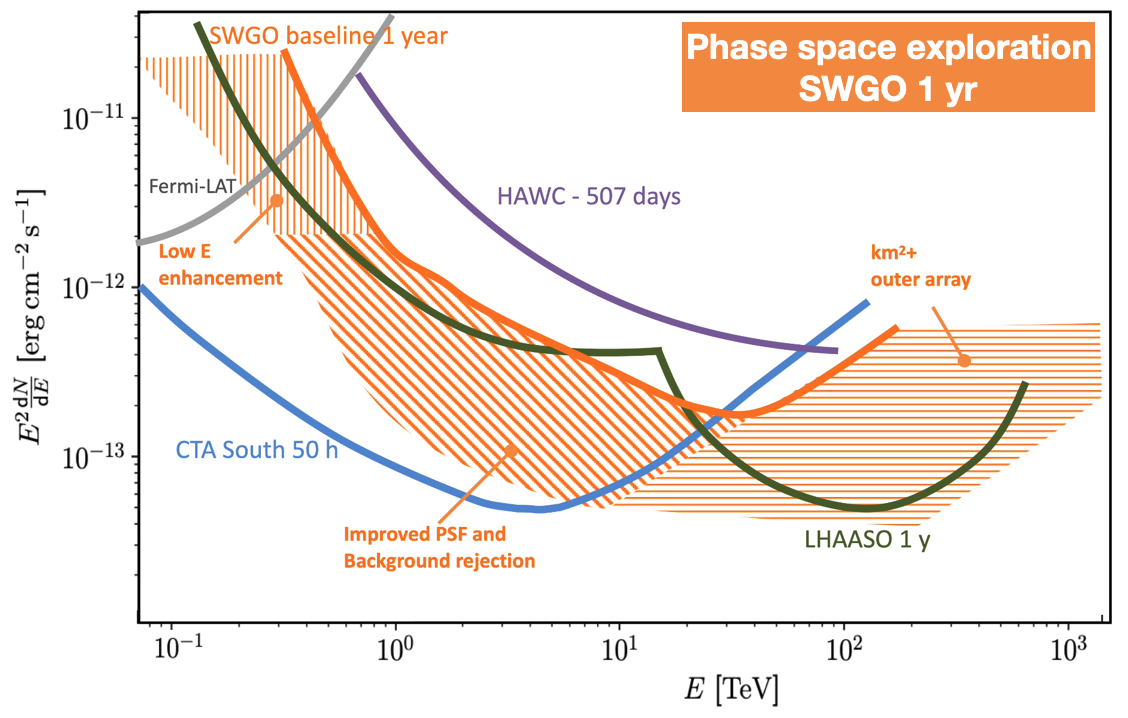}
  \caption{Differential point source sensitivity of several experiments (see labels) and phase-space exploration for SWGO. The orange bracketed phase-space is compared to the differential point-source sensitivity of various experiments. The \emph{baseline} curve represents the reference configuration. The lower limit of the orange band corresponds to a 30\% improvement in the point spread function (PSF) and a 10-fold enhancement in background rejection efficiency. The size of the outer array is the primary parameter driving the high-energy enhancement.
    \label{fig:Sensitivity}}
\end{figure}

While a definitive answer is not yet available, the ongoing research provides insights into the potential sensitivity achievable by SWGO, as indicated by the shaded area in Figure~\ref{fig:Sensitivity}.

In conclusion, SWGO is making steady progress despite challenges and demonstrates its potential as a powerful instrument in various domains, including very extended emission, transient phenomena, and beyond standard model physics searches. Collaborative efforts with CTA-South and LHAASO further enhances the scientific capabilities of SWGO, promising significant advancements in multi-messenger astronomy and full-sky coverage.

\section*{Acknowledgments}
The works developed at SWGO are supported by several funding agencies (see \url{https://www.swgo.org/SWGOWiki/doku.php?id=acknowledgements} for a complete list). This work has been financed by national funds through FCT - Fundação para a Ciência e a Tecnologia, I.P., under project PTDC/FIS-PAR/4300/2020.




\bibliographystyle{JHEP_no_title}
\bibliography{references}



\end{document}